\documentclass[lettersize,journal]{IEEEtran}
\usepackage{amsmath,amsfonts}
\usepackage{algorithmic}
\usepackage{algorithm}
\usepackage{array}
\usepackage[caption=false,font=normalsize,labelfont=sf,textfont=sf]{subfig}
\usepackage{textcomp}
\usepackage{stfloats}
\usepackage{url}
\usepackage{verbatim}
\usepackage{graphicx}
\usepackage{cite}
\begin{document}

\title{Optimisation of photodetectors design:\\comparison between Montecarlo\\ and Genetic Algorithms}

\author{Patricia M. E. V\'azquez\IEEEauthorrefmark{1}$^1$, Ligia Ciocci Brazzano\IEEEauthorrefmark{1}\IEEEauthorrefmark{2}, Francisco E. Veiras\IEEEauthorrefmark{1}, and Patricio A. Sorichetti\IEEEauthorrefmark{1}

\IEEEauthorblockA{\IEEEauthorrefmark{1}\emph{Universidad de Buenos Aires, Facultad de Ingenier\'ia, Departamento de F\'isica, GLOmAe.}\\\emph{Paseo Col\'on 850, C1063ACV, Buenos Aires, Argentina.}}

\IEEEauthorblockA{\IEEEauthorrefmark{2}\emph{Consejo Nacional de Investigaciones Cient\'ificas y T\'ecnicas de Argentina (CONICET).}\\\emph{Godoy Cruz 2290, C1425FQB, Buenos Aires, Argentina.}}

\IEEEauthorblockA{$^1$\texttt{\small{pvazquez@fi.uba.ar}}}}


\markboth{Journal of \LaTeX\ Class Files,~Vol.~14, No.~8, August~2021}%
{Shell \MakeLowercase{\textit{et al.}}: A Sample Article Using IEEEtran.cls for IEEE Journals}

\IEEEpubid{0000--0000/00\$00.00~\copyright~2021 IEEE}

\maketitle

\begin{abstract}
We present Montecarlo and Genetic Algorithm optimisations applied to the design of photodetectors based on a transimpedance amplifier and a photodiode. The circuit performance is evaluated with a merit function and the systematic search method is used as a reference. The design parameters are the feedback network components and the photodiode bias voltage. To evaluate the optimisations, we define the relative difference between its merit and the optimum merit obtained by the systematic search. In both algorithms, the relative difference decreases with the number of evaluations, following a power law. The power-law exponent for the Genetic Algorithm is larger than that of Montecarlo (0.74 vs. 0.50). We conclude that both algorithms are advantageous compared to the systematic search method, and that the Genetic Algorithm shows a better performance than Montecarlo. 
\end{abstract}


\section{Introduction}
\IEEEPARstart{O}{ptical} detectors based on transimpedance amplifiers (TIAs) \cite{Graeme1996} are widely used in arrays for measuring displacements \cite{Sorichetti2000}, in dynamic interferometry \cite{Riobo20172} \cite{Riobo2017}, in ultrasound generation \cite{Gonzalez2019}, in optoacoustics imaging and tomography \cite{Insabella2020bis} \cite{Insabella2020} \cite{Riobo2019}.

In the design of optical detectors, the selection of the design parameters values to comply with the specifications is very challenging \cite{Graeme1996} \cite{Vazquez2021} . A systematic search considering all the combinations of commercially available components will certainly lead to the optimum solution \cite{Vazquez2021}. However, this requires a large amount of calculations with significant computational cost. A popular approach to overcome this problem is trying a reasonable set of design parameters values with a circuit simulation tool, and verifying the performance. If the circuit does not comply with the specifications, we modify some parameters until a satisfactory result is obtained. Although this method is usually adequate for some applications, it does not necessarily lead to the optimum performance. For optical instrumentation designs, where some specifications (such as signal-to-noise ratio) might be quiet stringent, this approach is often not sufficient.

To accurately describe in a quantitative way the performances and compare them, merit functions are a useful tool. Merit functions are implemented in many engineering applications, such as calculations of gasoline compositions \cite{Ershov2022} \cite{Abdellatief2023}, lithographic lens \cite{Progler1996}, or thin films \cite{Dobrowolski1989}. 

Montecarlo (MC) is a statistical technique used to obtain approximate solutions to complex mathematical problems, including circuit optimisation. It consists in simulating a large number of random values of the design parameters. Then, the best result is selected with a specific criterion, for example, the higher value of a selected merit function. These MC methods are applied in very different fields, such as biomedical images \cite{Kyme2017}, and optoelectronic designs, for example, avalanche photodiodes \cite{Ma2002} \cite{LiewTatMun2008} \cite{Zheng2009}, and transimpedance amplifiers for optical communications \cite{Zohoori2018} \cite{SemsarParapari2021}. 

Genetic Algorithms (GA), used in optimisation problems, follow the steps of the biological evolution and natural selection \cite{HesamMahmoudiNezhad2014} \cite{Wirsansky2020} \cite{DiRado2014} \cite{Haupt2004} \cite{Thierens1994}. In this technique, the set of parameters values that define a circuit is called a chromosome. The algorithm begins with an initial population (first generation). Then, couples are formed, that are the parents of the next generation. The recombination and the natural selection process take place. A fraction of the chromosomes of each generation may be subjected to mutations. Genetic Algorithms are used for a variety of applications, such as the design of analog circuits \cite{Jiang2009} \cite{Cohen2015}, and integrated circuits \cite{Abdo2021} \cite{Devi2021} \cite{Settaluri2022}, materials characterization \cite{Paszkowicz2009} \cite{Boggi2014}, optical communication systems \cite{Kamalakis2021}, and LIDAR \cite{Muhire2022} \cite{Lee2022}. 

In this work, we present two different optimisations of the design of photodetectors based on transimpedance amplifiers: Montecarlo and Genetic Algorithm. The same merit function (defined in a previous work \cite{Vazquez2021}) is used for both optimisation techniques. The optimisation consists in finding the design parameters values that lead to the best photodetector performance, in terms of the constraints: signal-to-noise ratio, bandwidth and phase margin \cite{Vazquez2021} \cite{Kingston1978}. For this purpose, we use as a reference the systematic search presented in \cite{Vazquez2021}. We compare the results and the computational costs of the optimisations. 

This paper is structured as follows. In Section \ref{Specifications}, we present the specifications and definitions of the design that we optimise in this work. Section \ref{Section_GeneralDesignMethod} describes the general design method, including the optimisation algorithm, composed by the merit function and a search algorithm. In Section \ref{Section_Merit}, we describe the performance evaluation with the merit function. In Sections \ref{Montecarlo} and \ref{Geneticalgorithm}, the optimisations with the Montecarlo and with the Genetic Algorithm are introduced, respectively. In Section \ref{Section_Results}, we present the results and compare the performances of the two optimisation algorithms, using as a reference the systematic search. Finally, Section \ref{Section_Conclusions} summarizes the conclusions. 


\section{SPECIFICATIONS AND DESIGN DEFINITIONS}\label{Specifications}
\begin{figure}[ht]
\centering
\includegraphics[width=0.36\linewidth]{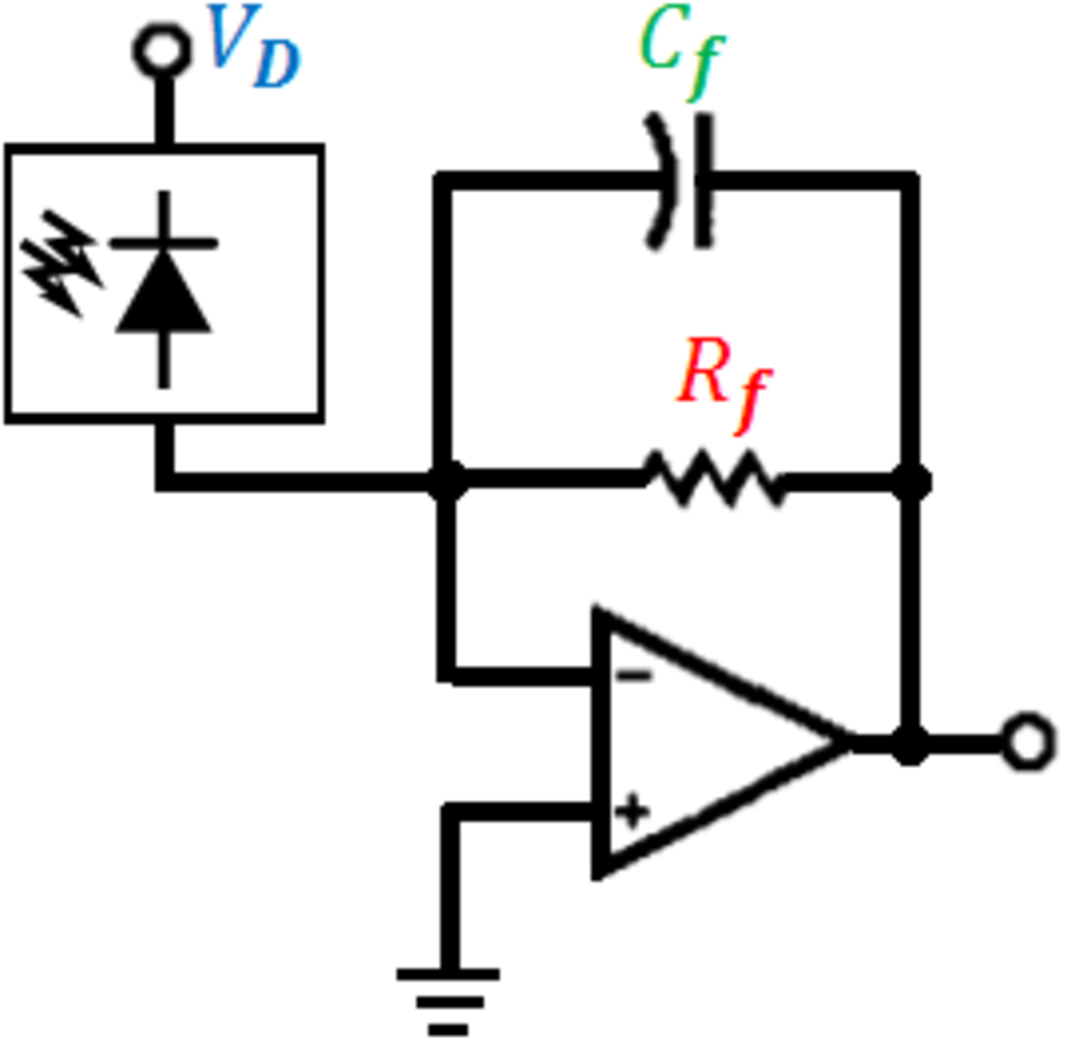}
\caption{Photodetector based on a transimpedance amplifier. The feedback network components are shown in red and green. The photodiode bias voltage node is indicated in blue.}
\label{fig:photodetector} 
\end{figure}

We propose an optical detector based on a photodiode and an operational amplifier in transimpedance configuration (Fig. \ref{fig:photodetector}).
 The small-signal model and the design equations, including the signal-to-noise ratio and stability analysis, are discussed in detail in a previous article \cite{Vazquez2021}.

In this work, the performance variables that define the specifications are the minimum signal-to-noise ratio (at minimum input irradiance), $S/N$; the highest operating frequencies, $B$; and the minimum phase margin value, $\phi_m$; and their tolerances. An additional constraint is that we only use commercial components that have standard values within a given tolerance (for example, $\pm10\%$, $\pm5\%$, etc).

The design parameters that we optimise are the feedback network components: the resistance $R_f$ and the capacitance $C_f$, and the photodiode bias voltage, $V_D$. Must be remarked that the photodiode capacitance, $C_D$, depends on the $V_D$ applied to the photodiode. This can be seen in the Diode Capacitance vs. Reverse Voltage plot of the photodiode, $C_D = f(V_D)$, from the manufacturers datasheet.

In Section \ref{Section_GeneralDesignMethod}, we explain the general design method of photodetectors, using a merit function and a search algorithm.

\section{GENERAL DESIGN METHOD}\label{Section_GeneralDesignMethod}
\begin{figure}[ht]
\centering
\includegraphics[width=0.9\linewidth]{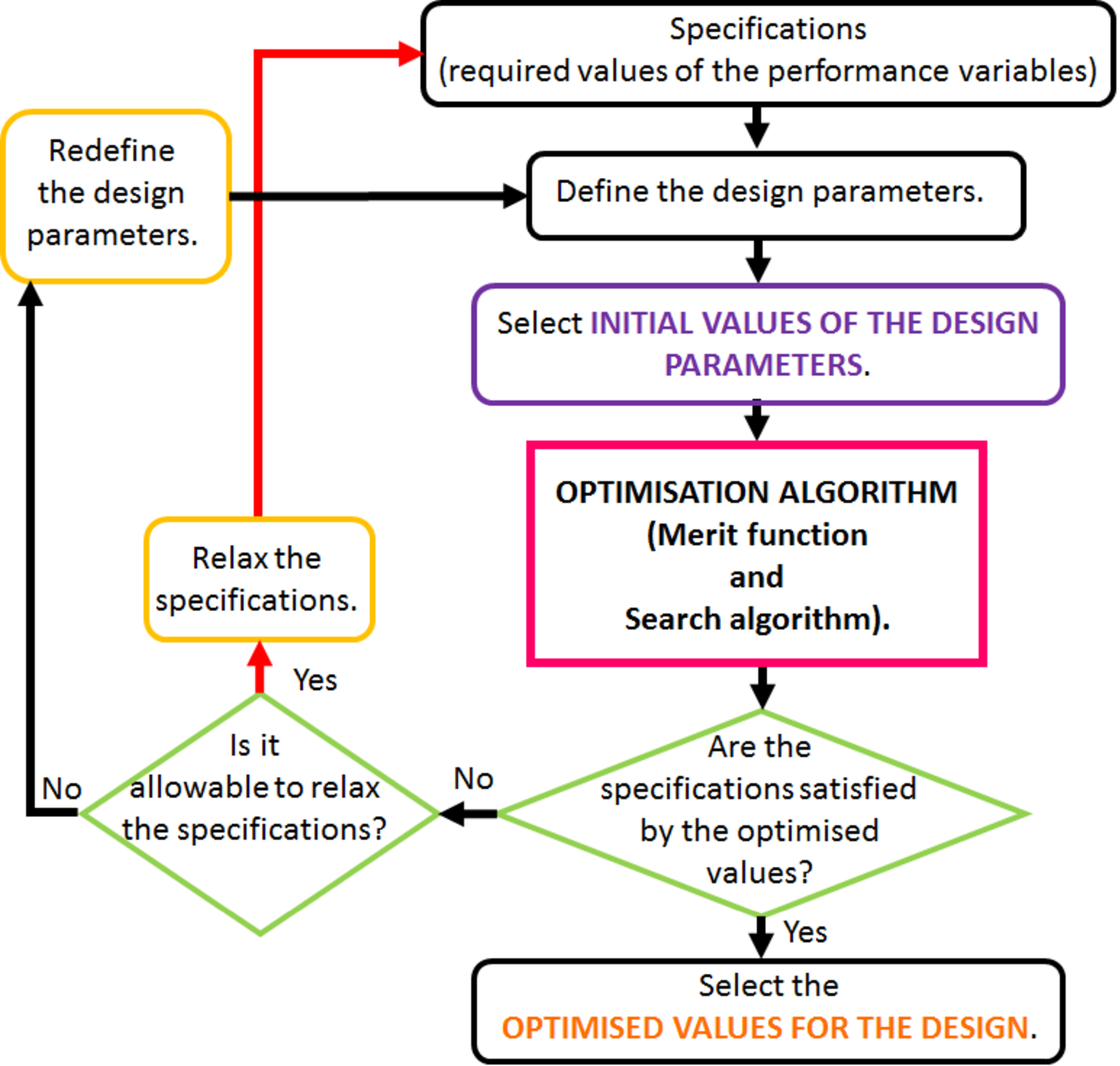}
\caption{Flow diagram of the general photodetectors design method.}
\label{fig:general_flow_diagram}
\end{figure}

We propose a general design methodology of photodetectors (Fig. \ref{fig:general_flow_diagram}). 
 We first take into consideration the specifications, i.e. the required values of the performance variables. We define the design parameters, whose values we optimise to satisfy the specifications. 

The first step is to set the initial values of the design parameters. Then, the optimisation algorithm (merit function and Search algorithm) is applied. The output of the optimisation algorithm gives the performance variables values with their merits. If the specifications are satified, we select the optimised design parameters values. On the contrary, if the specifications are not satisfied but they can be relaxed, a new optimisation can be carried out. If it were not possible to relax the specifications, the design parameters must be redefine.

In the next section, we present the merit function to evaluate the performance variables values. 

\section{MERIT FUNCTION}\label{Section_Merit}
We evaluate the performance achieved by a given set of design parameters values with the following global merit function, introduced in a previous work \cite{Vazquez2021}:
  \begin{equation}
Merit = M(x_1)M(x_2)...M(x_N),
\label{eq1}
\end{equation}
where $Merit$ indicates the global degree of compliance with the specifications, and $M(x_i)$ is the merit corresponding to the performance variable, $x_i$. We calculate merit evaluating the values of the performance variables that define the specifications; in this work, $S/N$, $B$ and $\phi_m$. The latter are obtained as functions of the design parameters ($R_f$, $C_f$, and $V_D$) \cite{Vazquez2021}. For each performance variable, the specification defines an optimum value, $x_{opt}$, and its tolerance defines an acceptable region.

 \begin{figure}[ht]
\centering
\includegraphics[width=0.55\linewidth]{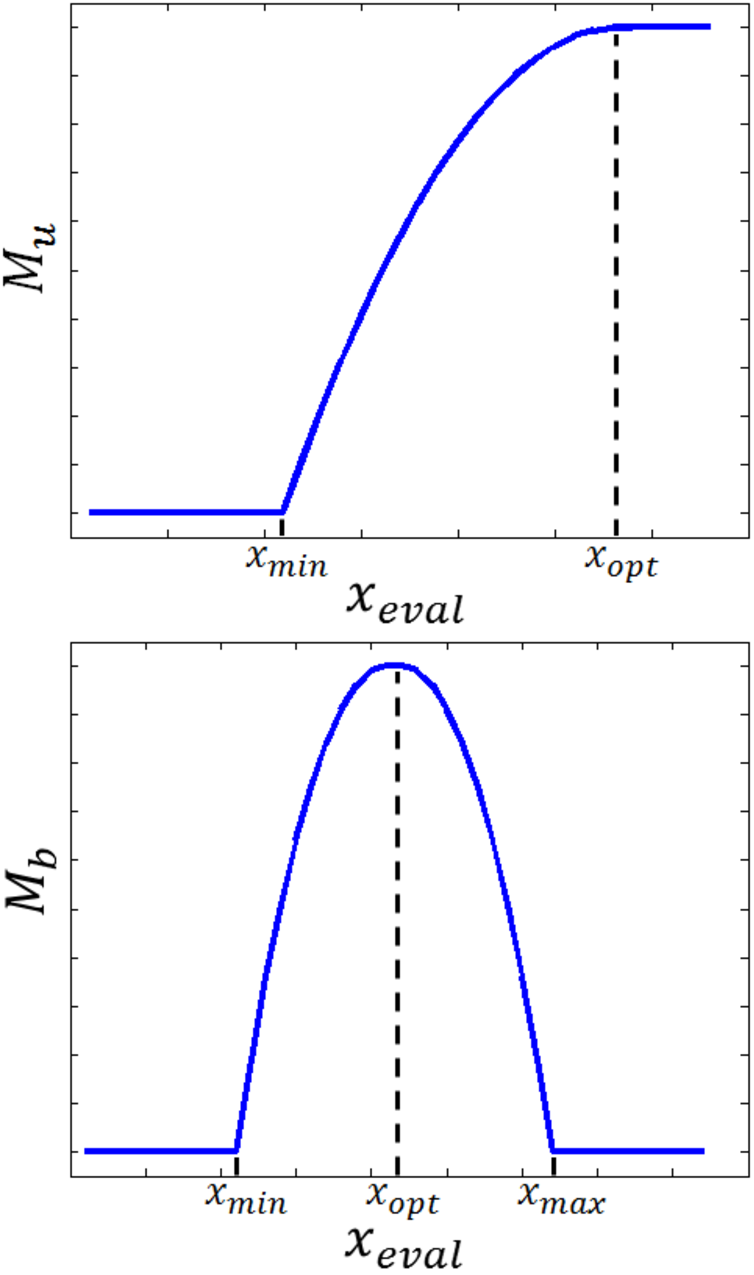}
\caption{Upper panel: unilateral merit function; lower panel: bilateral merit function. We choose these functions to weight the performance variables ($x$).}
\label{fig:merits} 
\end{figure}

If the performance variable, $x$, has an acceptable region limited by a lower bound, $x_{min}$, and an optimum, $x_{opt}$, above which a higher value has no advantage, we propose the unilateral merit function (Fig. 3, upper panel), as follows:
\begin{equation}
M_u(x) = \begin{cases} 0  & \text{for } x \leq x_{min}, \\ 1-\big(\frac{x-x_{opt}}{x_{min}-x_{opt}}\big)^2 & \text{for } x_{min} \leq x < x_{opt},\\ 1 & \text{for } x_{opt} \leq x.\end{cases}
\label{Mu}
\end{equation}
In this work, the merits of $S/N$ and $\phi_m$ are defined by the Unilateral merit functions, $M_{uS/N}$ and $M_{u\phi_m}$, respectively. It is easy to see that an analogous unilateral merit function may be defined if the acceptable region has an upper bound.

On the other hand, if the performance variable, $x_{eval}$, has an acceptable region limited by lower and upper bounds, $x_{min}$ and $x_{max}$, we use the Bilateral merit function (Fig. 3, lower panel):
   \begin{equation}
M_b(x) = \begin{cases} 0  & \text{for } x \leq x_{min},  \\  1-\big(\frac{x-x_{opt}}{x_{min}-x_{opt}}\big)^2  & \text{for } x_{min} \leq x < x_{opt}, \\ 1-\big(\frac{x-x_{opt}}{x_{max}-x_{opt}}\big)^2 & \text{for } x_{opt} \leq x < x_{max},\\ 0  & \text{for } x_{max} \leq x.
\end{cases}
\label{Mb}
\end{equation}
In this work, the tolerance of $B$ defines the lower and upper bounds of its acceptable region; therefore, the bilateral merit function is $M_{bB}$.

In the following sections, we introduce algorithms to search for the optimum merit.

\section{OPTIMISATION WITH MONTECARLO ALGORITHM}\label{Montecarlo}
\begin{figure}[ht]
\centering
\includegraphics[width=0.95\linewidth]{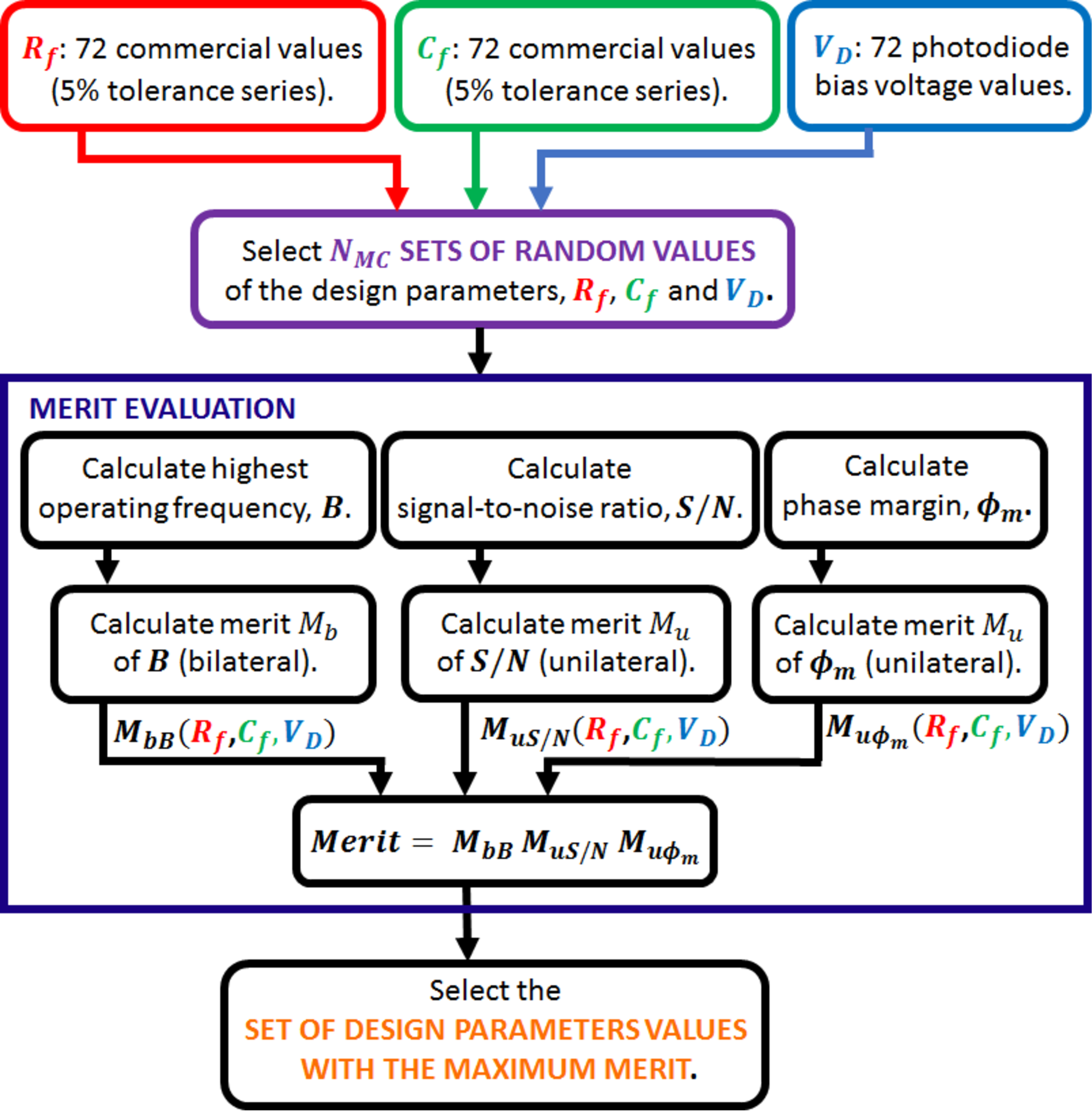}
\caption{Flow diagram of the Montecarlo algorithm.}
\label{fig:diagrama_flujo_MonteCarlo} 
\end{figure}

In this section, we present a Montecarlo algorithm to find the optimum photodetector design parameters values (Fig. \ref{fig:diagrama_flujo_MonteCarlo}). 

As defined above, the design parameters are the feedback network components, $R_f$ and $C_f$, and the photodiode bias voltage, $V_D$. Another constraint of our photodetector design is that the components must have commercial values.

We select the $72$ commercial values from the $5\%$ tolerance series for $R_f$ and $C_f$. For the bias voltage values, we discretize the photodiode reverse voltage  range (from the manufacturer datasheet) into $72$ values.

From the space of
\begin{equation}
    N_{3DP} = 72\ \times 72\ \times72 \approx 3.7 \times 10^5
    \label{eq4}
\end{equation} design parameters values, the algorithm is initialized with $N_{MC}$ sets of random design parameters values (each one with a $R_f$, a $C_f$, and a $V_D$ values). For each set of design parameters values, we calculate the performance variables: $B$, $S/N$, and $\phi_m$ (these calculations are detailed in \cite{Vazquez2021}). Then, we calculate the Unilateral merits corresponding to $S/N$ and $\phi_m$, $M_{uS/N}$ and $M_{u\phi_m}$, as in Eq. (\ref{Mu}); and the Bilateral merit of $B$, $M_{bB}$, as in Eq. (\ref{Mb}). The global $Merit$ is calculated as in Eq. (\ref{eq1}). Finally, we select the set of design parameters values with the maximum global merit.

In Section \ref{Geneticalgorithm}, we introduce a Genetic Algorithm for the design optimisation of photodetectors based on transimpedance amplifiers, using the same merit function.

\section{OPTIMISATION WITH GENETIC ALGORITHM}\label{Geneticalgorithm}
\begin{figure}[ht]
\centering
\includegraphics[width=0.85\linewidth]{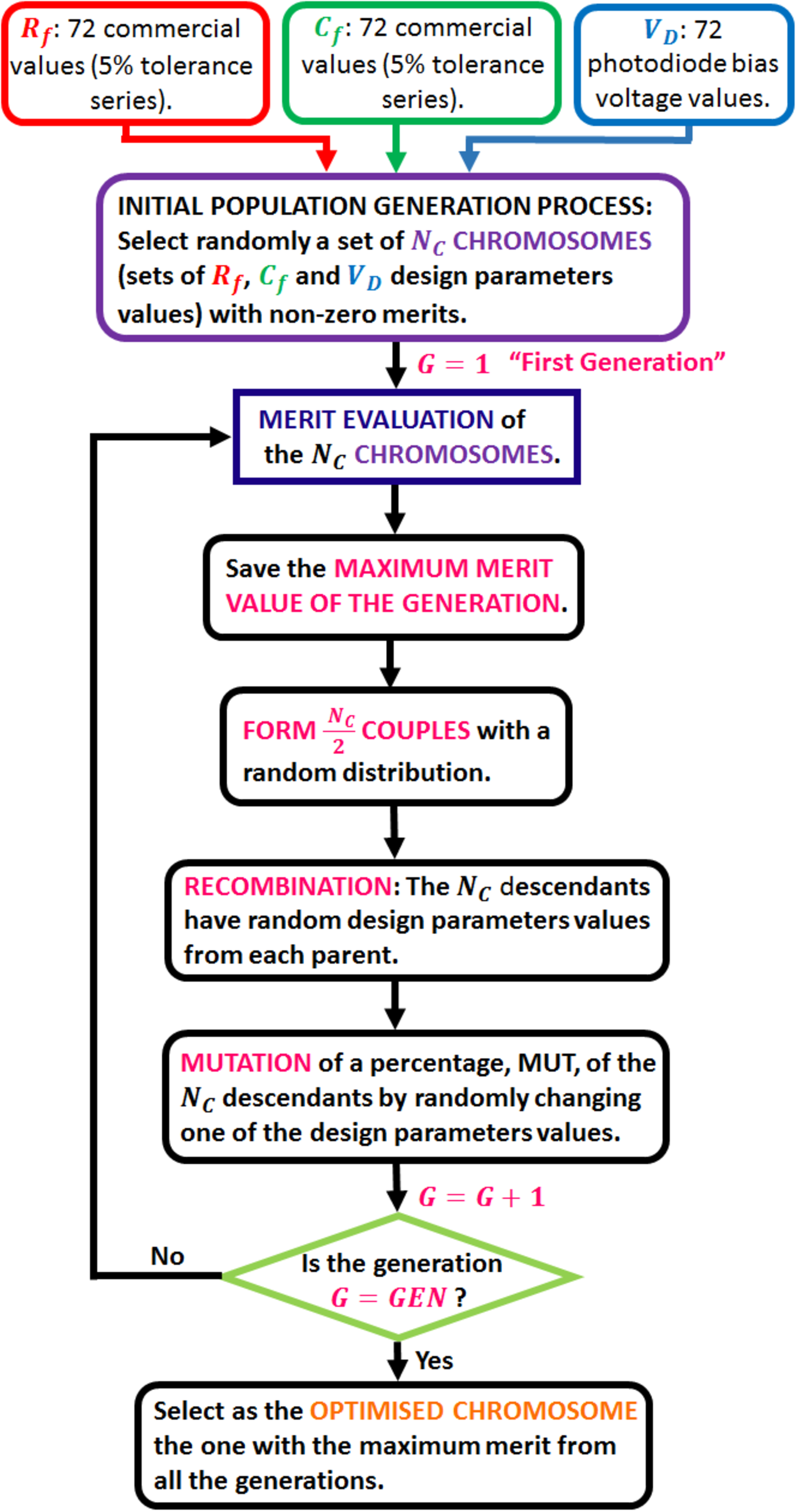}
\caption{Flow diagram of the optimisation with the Genetic Algorithm, initialized with $N_{C}$ chromosomes.}
\label{fig:GA_flow_diagram}
\end{figure}

We present a Genetic Algorithm to optimise the search of the design parameters values (Fig. \ref{fig:GA_flow_diagram}) for an optical detector based on a photodiode and a transimpedance amplifier (Fig. \ref{fig:photodetector}).
To compare the performance of the Genetic Algorithm optimisation with that of Montecarlo, we use the same design example and specifications of the previous section. Thus, the design parameters are $R_f$, $C_f$ and $V_D$. For the feedback network components, we select $70$ $R_f$ commercial values and $27$ $C_f$ commercial values from the $5\%$ tolerance series, and the $V_D$ are $70$ discrete values from the photodiode operation range. This is the same set of values that was used for the optimisation with Montecarlo algorithm.

We initialize the algorithm by randomly choosing the First Generation of Chromosomes \cite{DiRado2014}. For this purpose, we randomly select design parameters values that lead to $N_C$ circuits with non-zero merit. 

Then, Natural Selection takes place choosing which chromosomes are apt to survive. The merit evaluation of the chromosomes is carried out, using the merit function introduced in Section \ref{Section_Merit}. In our design, couples (the parents of the next generation) are formed with a proportional selection method \cite{DiRado2014} \cite{Thierens1994} \cite{Grefenstette1997}. First, a chromosome is randomly selected. Then, the chromosome can be chosen as a parent with a probability proportional to its merit value. This method allows to choose the best individuals with a higher probability, and also allows chromosomes with zero merit to be chosen. This helps maintaining the diversity of the population. In our selection method, a parent can be selected for more than one couple. The total amount of couples is $N_{C}/2$.

Each couple has two descendants. For this purpose, the recombination occurs, so that descendants contain a random amount of genetic content (in our case, design parameters values) from each one of the two parents. Then, considering that in this example there are three design parameters, recombination can take place in three different modes (A, B or C), as shown in this example: 
 \begin{equation*}
 \centering
  \textbf{Couple 1}\begin{cases}
 \textbf{Parent 1:}\ R_{f1} C_{f1} V_{D1} \ |\ 
  \textbf{Parent 2:}\ R_{f2} C_{f2} V_{D2}.
  \end{cases}
  \end{equation*}
  \begin{equation}
      \centering
       \begin{cases}\ \ \ \ \textbf{Descendant 1}\ \ | \ \ \textbf{Descendant 2}\\ \  \textbf{(A)}\ R_{f1}  C_{f2}  V_{D2}\ \ |\ \ R_{f2}  C_{f1}  V_{D1},\\ \ \textbf{(B)}\ R_{f1}  C_{f1}  V_{D2} \ \ | \ \ R_{f2}  C_{f2}  V_{D1},\\ \ \textbf{(C)}\ R_{f1}  C_{f2}  V_{D1} \ \ |\ \ R_{f2}  C_{f1}  V_{D2}.\end{cases}
     \label{eq5}
 \end{equation}
This means that each couple (such as Couple 1, formed by Parent 1 and Parent 2) reproduces into two descendants from the next generation (Descendant 1 and Descendant 2), by randomly recombining as in modes (A), (B), or (C) (Eq. (\ref{eq5})). Therefore, we recombine the chromosomes from the selected couples, leading to the creation of the Second Generation. 

To avoid the convergence to local maxima and to increase the diversity, mutations are introduced. We aim to find the maximum merit possible in our search of the optimised design parameters values (the optimised chromosome). Mutations are random changes of the design parameters values, that are applied to a percentage of chromosomes, $MUT$, of our descendants generation (from the Second Generation onwards). 

We preset the maximum number of generations, $GEN$.  When the Genetic Algorithm reaches the generation number $GEN$, it stops and we select as the optimised chromosome the one with the maximum merit from all the generations.

In the next section, we show the results obtained from the two different optimisations: Montecarlo algorithm and the Genetic Algorithm.

\section{RESULTS}\label{Section_Results}
We compare the performance of the Montecarlo algorithm and the Genetic Algorithm, using as a reference the systematic search presented in a previous work \cite{Vazquez2021}. We use these algorithms to search for the optimised design parameters values for optical detectors based on a photodiode and a transimpedance amplifier (Fig. \ref{fig:photodetector}). We present a design example to compare the three search algorithms with the definitions from Section \ref{Specifications}, and the following specifications:
\begin{equation}
\begin{cases}
S/N > 10\ \text{dB}, \\
B = (22 \pm 2)\ \text{kHz}, \\
  \phi_m > 45^{\circ}.  
\end{cases}
\end{equation}
A constraint is that we only use commercial components from the $5\%$ tolerance series. We choose to work with a PIN photodiode, BPW34 \cite{BPW34}, and a low-noise operational amplifier, OP07 \cite{OP07}. The design parameters are the feedback network components, $R_f$ and $C_f$, and the photodiode bias voltage, $V_D$.
In a previous work \cite{Vazquez2021}, we detailed this same photodetector small-signal model, and its noise and stability analysis. In that work, we introduced a merit function, that we also use in this work, and the systematic search method of the optimum design parameters. In \cite{Vazquez2021}, the design parameters were the feedback network components values, considering all the possible combinations of the $R_f$ and $C_f$ commercial values ($5\%$ tolerance series). It must be remarked that the specifications in \cite{Vazquez2021} are the same as in this work.

First, we use the systematic search method to determine the optimum design parameters values of the photodetector. Considering that there are $72$ commercial components values in the $5\%$ tolerance series, we discretize the allowable $V_D$ values of the photodiode into $72$ values. Then, all the possible combinations of design parameters values, $R_f$, $C_f$ and $V_D$, in this case, lead to $N_{3DP} \approx 3.7 \times 10^5$ computational evaluations (Eq. \ref{eq4}).
The optimum design parameters of the systematic search are
  \begin{equation}
      \centering
       \begin{cases}
       R_f = 620\ $k$\Omega,\\ 
       C_f = 3.6\ $pF$,\\
       V_D = 14.79\ $V,$
\end{cases} 
     \label{eq7}
 \end{equation}
 and the resulting performance values are
  \begin{equation}
      \centering
       \begin{cases}
       Merit_{syst} = 0.9338,\\ 
       S/N = 71\ $dB,$\\
       B = 22.01\ $kHz,$\\
       \phi_m = 90^{\circ}.\\
\end{cases}
     \label{eq8}
 \end{equation}The merits corresponding to the performance variables are
  \begin{equation}
      \centering
       \begin{cases}
       M(S/N) = 0.9338,\\ 
       M(B) = 1,\\
       M(\phi_m) = 1.\\
\end{cases}
     \label{eq9}
 \end{equation}
The computational time of the systematic search is $t = 27.1$ s. Since there are three design parameters, the domain of the merit function, $Merit_{syst}$, is 3D. Then, to visualize the results, we plot 2D projections of the merit function. \begin{figure}[ht]
\centering
\includegraphics[width=1\linewidth]{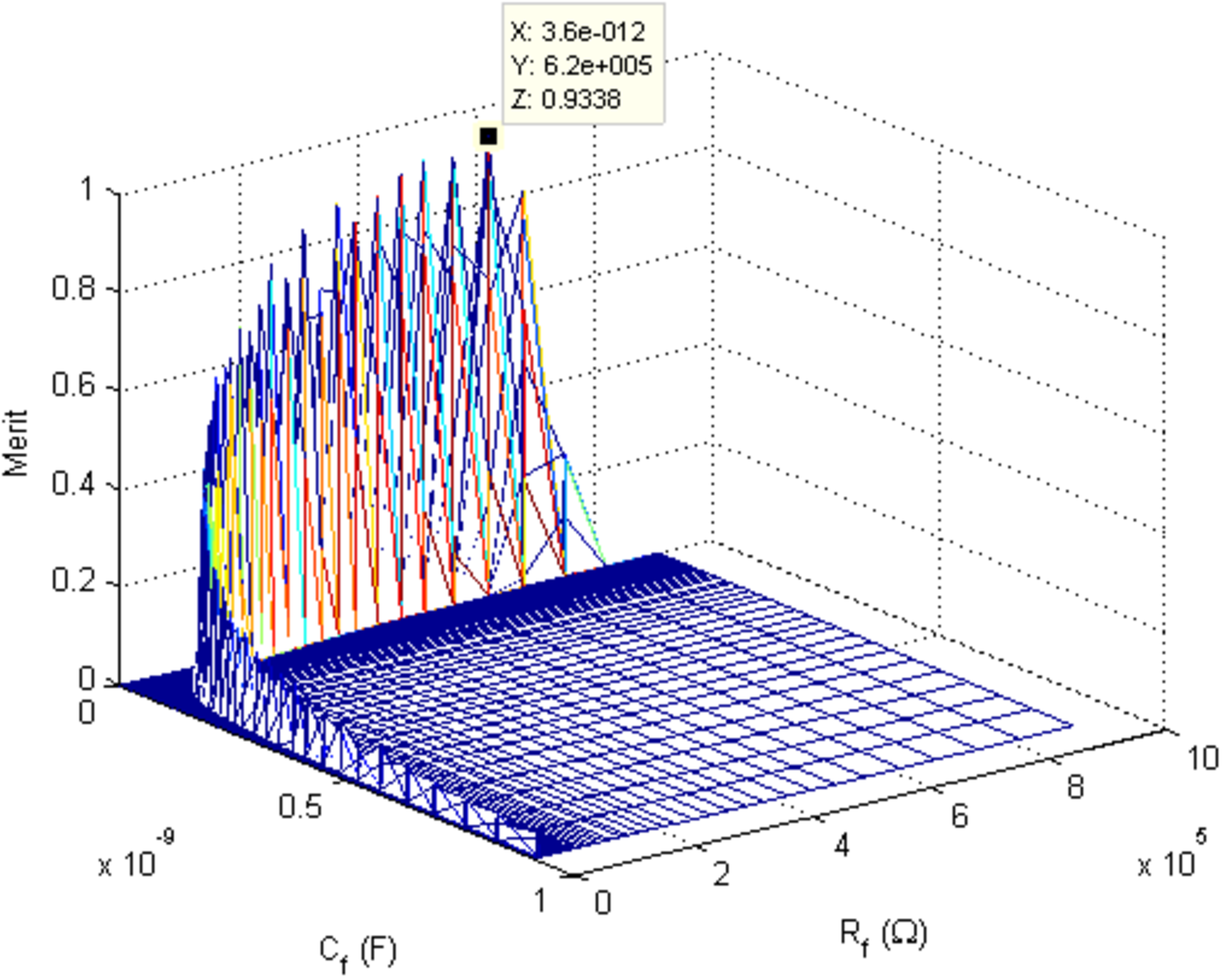}
\caption{Systematic search merits as a function of the feedback network components, $C_f$ and $R_f$, evaluated at the optimum $V_D$ value. The optimum global merit value, $Merit_{syst}$, is indicated with a black square.}
\label{fig:Vdcte}
\end{figure}
\begin{figure}[ht]
\centering
\includegraphics[width=1\linewidth]{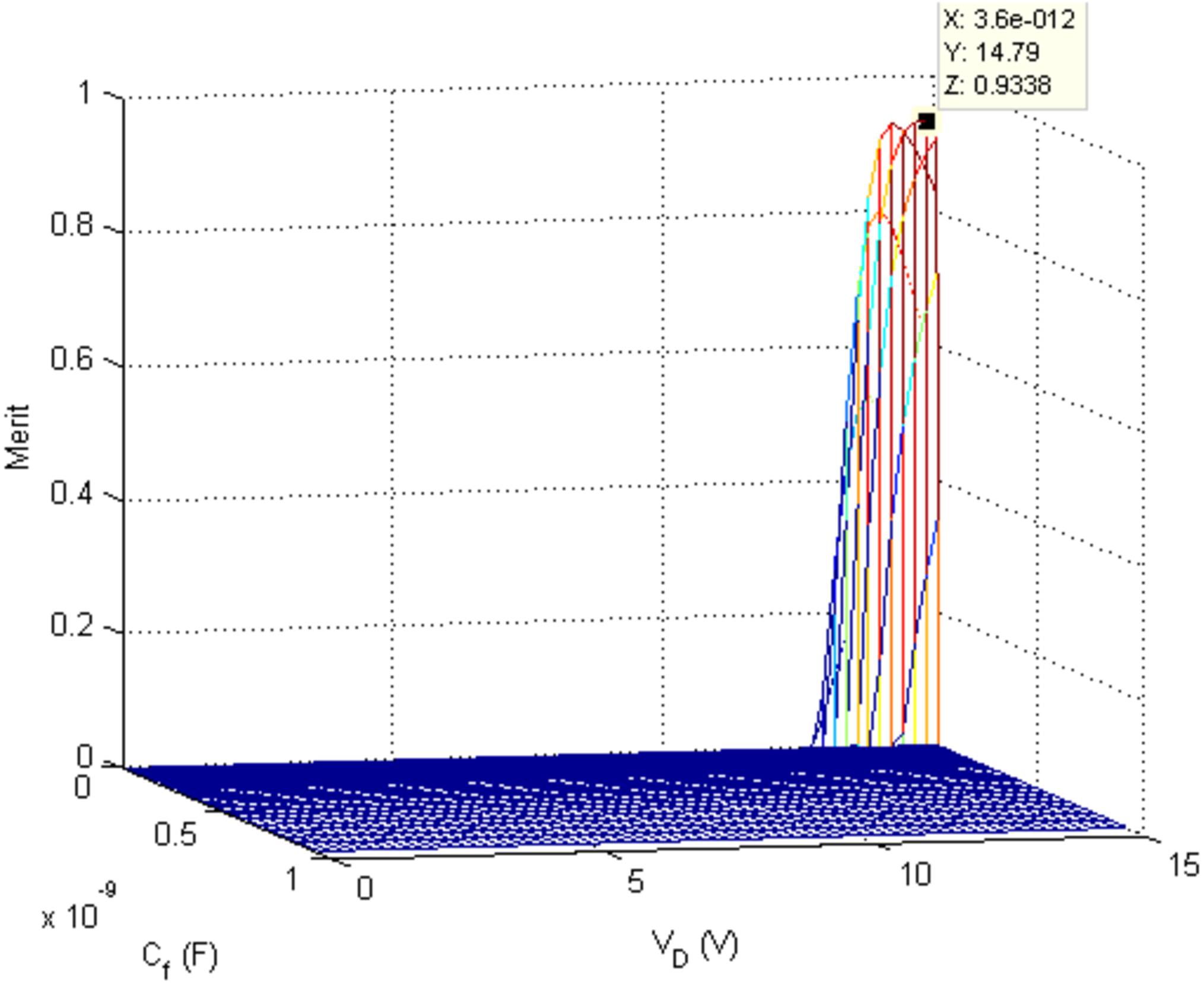}
\caption{Systematic search merits as a function of the feedback capacitance, $C_f$, and the photodiode bias voltage, $V_D$, evaluated at the optimum  $R_f$ value. The optimum global merit value, $Merit_{syst}$, is indicated with a black square.}
\label{fig:Rfcte}
\end{figure}
\begin{figure}[ht]
\centering
\includegraphics[width=1\linewidth]{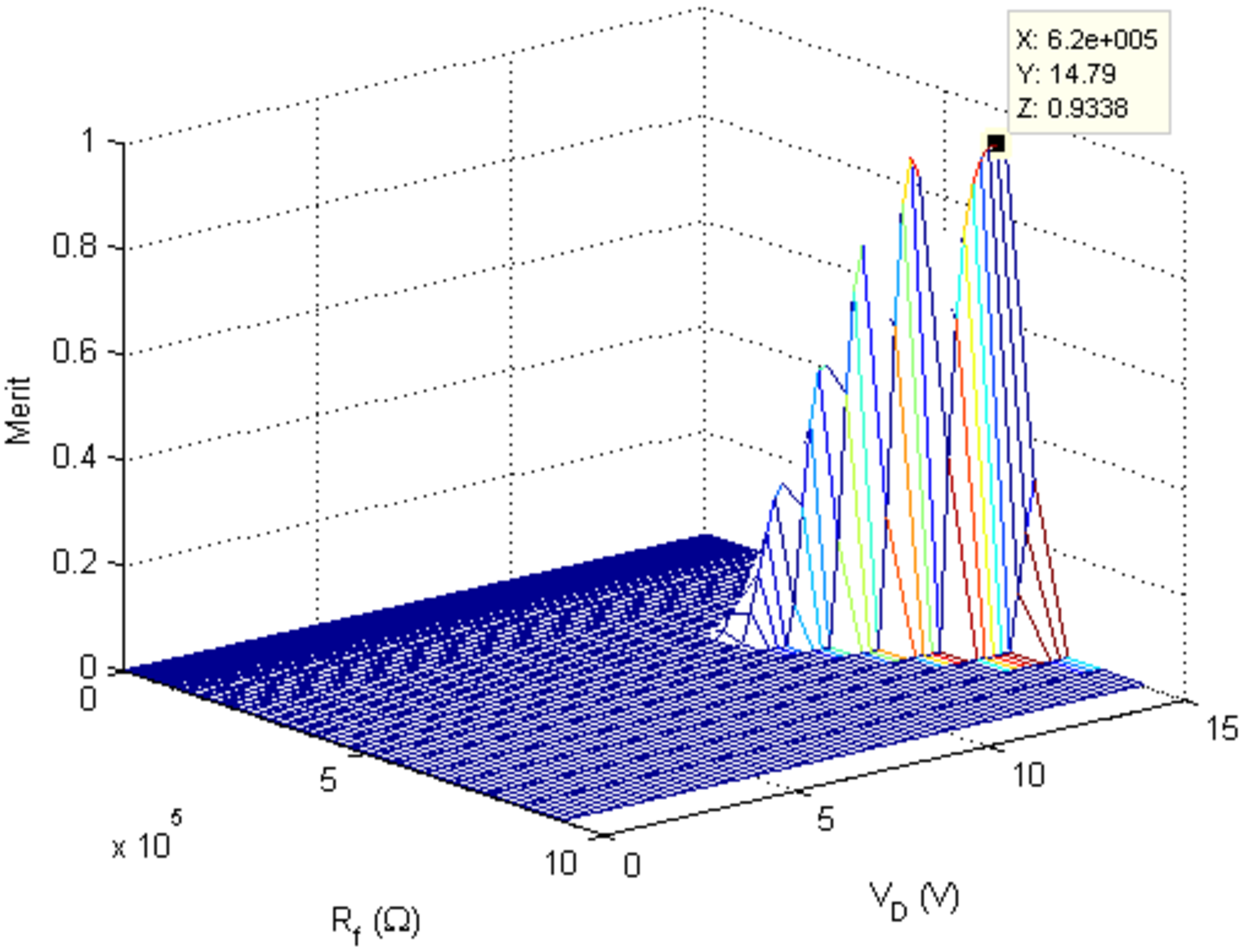}
\caption{Systematic search merits as a function of the feedback network resistance, $R_f$, and the photodiode bias voltage, $V_D$, evaluated at the optimum  $C_f$ value. The optimum global merit value, $Merit_{syst}$, is indicated with a black square.}
\label{fig:Cfcte}
\end{figure} For each projection,  we consider the optimum value of one of the design parameters and we plot the merit as a function of the other two (Figs. \ref{fig:Vdcte}, \ref{fig:Rfcte} and \ref{fig:Cfcte}). In addition, the optimum global merit value, $Merit_{syst}$, is indicated. From the figures, we see that the optimum lies within a narrow, well-defined region. 

From these results, we conclude that including $V_D$ as a design parameter significantly improves the photodetector performance compared to \cite{Vazquez2021}, where $R_f$ and $C_f$ were the only design parameters. $Merit_{syst}$ increases from 0.87 to 0.93, and the signal-to-noise ratio improves from $31$ dB to $71$ dB.

 \begin{figure*}[h]
\centering
\includegraphics[width=1\linewidth]{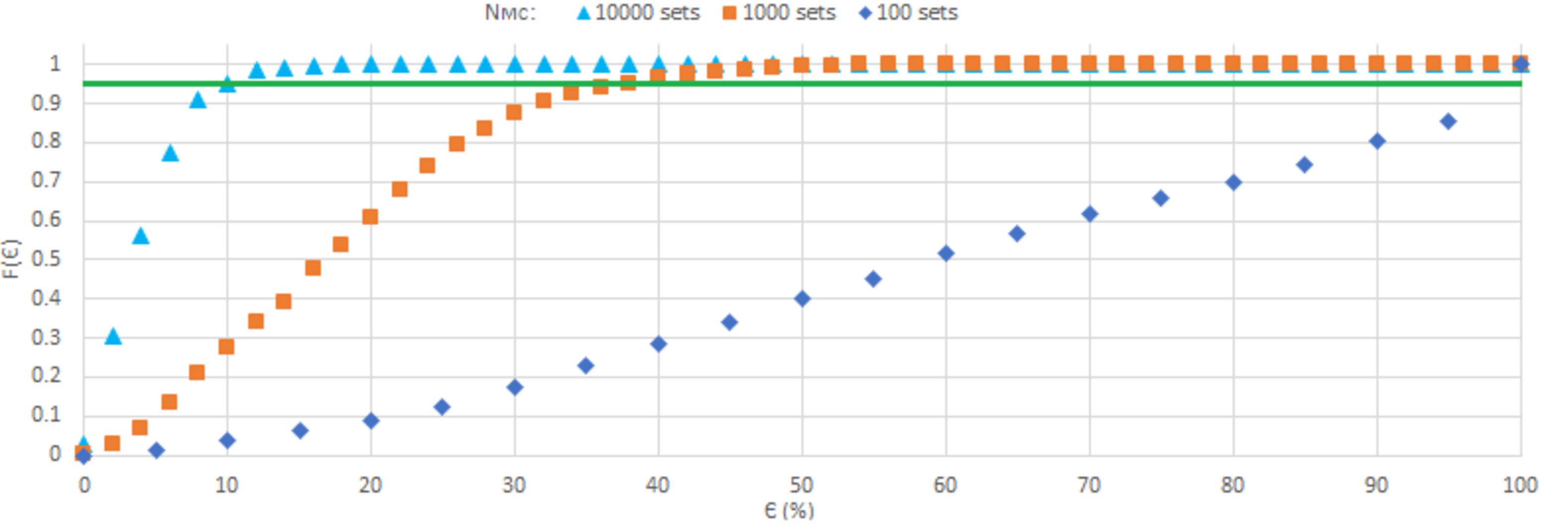}
\caption{Normalized cumulative distribution, $F(\epsilon)$, of the percent relative differences, $\epsilon$, between $Merit_{syst}$ and the Montecarlo optimised result, $Merit$. Each plot corresponds to $1000$ MC experiments with different numbers of sets of random design parameters values, $N_{MC}$. The green line indicates $F(\epsilon) = 0.95$.}
\label{fig:MC_histograms} 
\end{figure*}

We define $\epsilon$, the percent relative difference between the optimum merit obtained by the systematic search, $Merit_{syst}$, and the optimised result, $Merit$, applying the optimisation algorithm (in this work, Montecarlo or Genetic Algorithm):
\begin{equation}
\epsilon = 100\frac{Merit_{syst}-Merit}{Merit_{syst}},\ \ 0\% < \epsilon < 100\%;
\label{eq10}
\end{equation}
clearly, $Merit$ and, therefore, $\epsilon$ are random variables. A useful parameter is $\epsilon_{95}$, defined by:
\begin{equation}
P(\epsilon<\epsilon_{95})= 95\%,
\label{eq11}
\end{equation} that is to say, in $95\%$ of the experiments, $\epsilon$ is lower than $\epsilon_{95}$.

We run $1000$ times the Montecarlo search algorithm to carry out a statistical analysis of the results. Each run of the Montecarlo search algorithm is a MC experiment. In each MC experiment are evaluated $N_{MC}$ sets of random values of the design parameters (Fig. \ref{fig:diagrama_flujo_MonteCarlo}).

\begin{figure}[ht]
\centering
\includegraphics[width=1\linewidth]{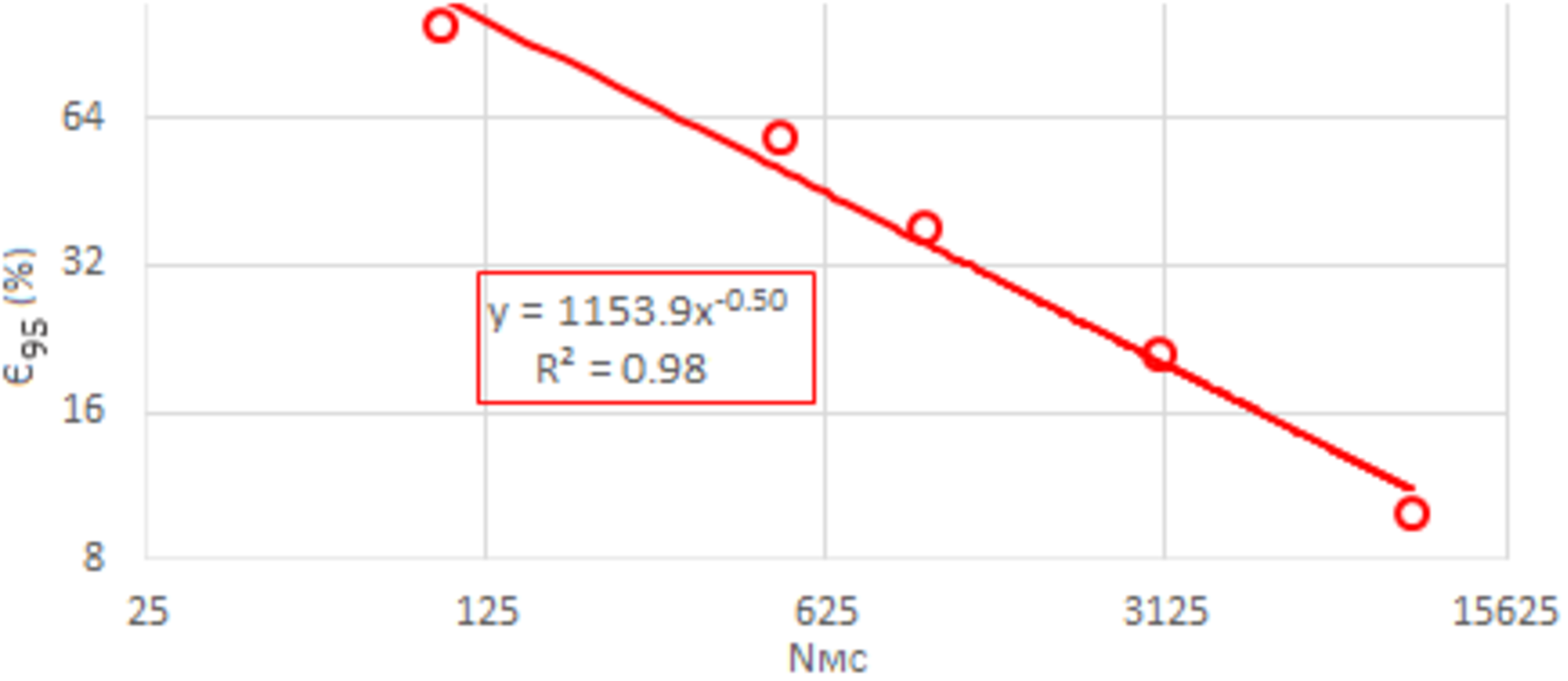}
\caption{$\epsilon_{95}$ as a function of the number of sets of random design parameters values, $N_{MC}$.}
\label{fig:E95_Nmc}
\end{figure}

In Fig. \ref{fig:MC_histograms}, we show the normalized cumulative distribution, $F(\epsilon)$, corresponding to 1000 Montecarlo experiments, with different values of $N_{MC}$. 
\begin{table}[ht]
\caption{\label{tab1}Montecarlo search algorithm results.}
\begin{center}
    \begin{tabular}{cccccccc}
      $N_{MC}$ & $\epsilon_{95}$ (\%) & $t$ (s)\\
         \hline
      $100$  &  $> 99$ & $0.014$\\ 
      $500$  &  $58$ & $0.043$\\ 
      $1000$  &  $38$ & $0.080$\\ 
      $3000$  &  $21$ & $0.24$\\ 
      $10000$  &  $10$ & $0.72$\\ 
\end{tabular}
\end{center}
\end{table} 
Table \ref{tab1} summarizes the $\epsilon_{95}$ values and computation times for each $N_{MC}$. We see, in Fig. \ref{fig:E95_Nmc},
that $\epsilon_{95}$ follows a power law with $N_{MC}$, that is
\begin{equation}
\epsilon_{95} \propto N_{MC}^{-\beta}, 
\label{eq12}
\end{equation}with $\beta \approx 0.50$, while $t$ linearly increases. For example, in a MC experiment with $10000$ random sets of design parameters, $\epsilon_{95}$ is $10\%$. Clearly, the result is close to the optimum merit value from the systematic search, with $37$ times less evaluations ($10^4$ vs. $N_{3DP} \approx 3.7 \times 10^5$).

 \begin{figure*}[h]
\centering
\includegraphics[width=1\linewidth]{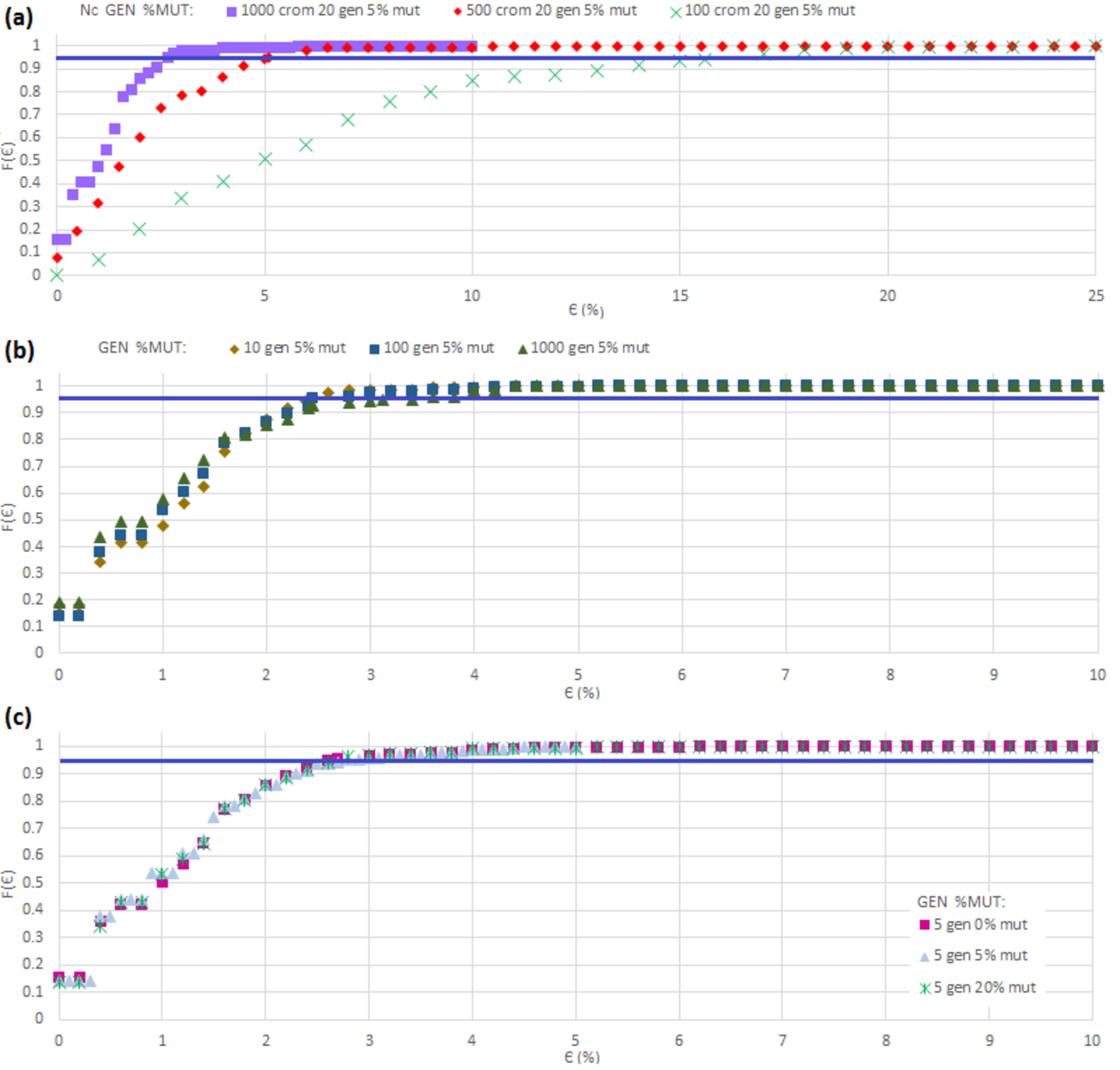}
\caption{Normalized cumulative distribution, $F(\epsilon)$, of the percent relative differences, $\epsilon$, between $Merit_{syst}$ and the Genetic Algorithm optimised chromosome result, $Merit$. The blue line indicates $F(\epsilon) = 0.95$. (a) Each plot corresponds to $200$ GA experiments with different numbers of chromosomes ($N_{C}$), $20$ generations, and $5\%$ mutations. (b) Each plot corresponds to $1000$ GA experiments with $1000$ chromosomes, different numbers of generations ($GEN$), and $5\%$ mutations. (c) Each plot corresponds to $1000$ GA experiments with $1000$ chromosomes, $5$ generations, and different percentages of mutations, $MUT$.}
\label{fig:GA_acum}  
\end{figure*}

The three numerical parameters that determine the performance of the Genetic Algorithm are the number of chromosomes, $N_C$; the number of generations, $GEN$; and the percentage of mutations, $MUT$ (cf. the flow diagram in Fig. \ref{fig:GA_flow_diagram}). A GA experiment is a run of the Genetic Algorithm, for fixed values of $N_C$, $GEN$ and $MUT$. The number of circuit evaluations in each GA experiment, $Eval_{GA}$, is useful to measure its computational cost. For the Genetic Algorithm,
\begin{equation}
Eval_{GA} = GEN \cdot N_C.
\label{eq13}
\end{equation}
This allows the comparison with other algorithms. For instance, the systematic search algorithm requires $N_{3DP} \approx 3.7 \times 10^5$ circuit evaluations. 

To evaluate the algorithm performance, we determine the $\epsilon_{95}$ for different values of of $N_C$, $GEN$ and $MUT$. In Fig. \ref{fig:GA_acum} (a), we plot the normalized cumulative distribution, $F(\epsilon)$ (Eq. \ref{eq10}). Results are shown from $200$ GA experiments, with different $N_{C}$ values ($100$, $500$ and $1000$), $20$ generations, and $5\%$ mutations. In this figure, we can see that the $\epsilon_{95}$ clearly improves with increasing $N_C$. 
\begin{table}[ht]
\caption{\label{tab2}Genetic Algorithm results.}
\begin{center}
\begin{tabular}{cccccccc}
 $N_{C}$ & $GEN$ &  $MUT\ (\%)$ & $Eval_{GA}$ & $\epsilon_{95} (\%)$ & $t$ (s)\\    \hline
 $100$ & $20$&  $5$ & $2000$ & $15.6$  &$1.0$\\ 
  $500$ & $20$ & $5$& $10000$ & $5.10$  &$4.6$\\ 
 $1000$ & $20$ & $5$ & $20000$& $2.69$  &$9.1$\\  \hline
  $1000$ & $10$& $5$ & $10000$ &$2.38$   &$8.2$\\ 
  $1000$ & $100$ & $5$ & $100000$ & $2.44$   &$15.0$\\ 
  $1000$ & $1000$ & $5$ & $1000000$&$3.12$   &$88.0$\\ \hline 
$1000$ & $5$ & $0$ & $5000$& $2.44$  &$7.6$\\ 
$1000$ & $5$ & $5$ & $5000$ &$2.70$  &$7.5$\\ 
$1000$ & $5$ & $20$ & $5000$ &$2.69$  &$7.5$\\  
\end{tabular}
\end{center}
\end{table} 
These results are shown in the first three rows of Table \ref{tab2}, where we summarize the GA results. From this table follows that $Eval_{GA}$ and the computation time of the GA experiment, $t$, increase with $N_C$. 

Fig. \ref{fig:GA_acum} (b) shows that $\epsilon_{95}$ does not appreciably change when the $GEN$ value goes from $10$ to $1000$, for $N_C = 1000$ and $MUT = 5\%$. \begin{figure}[ht]
\centering
\includegraphics[width=0.95\linewidth]{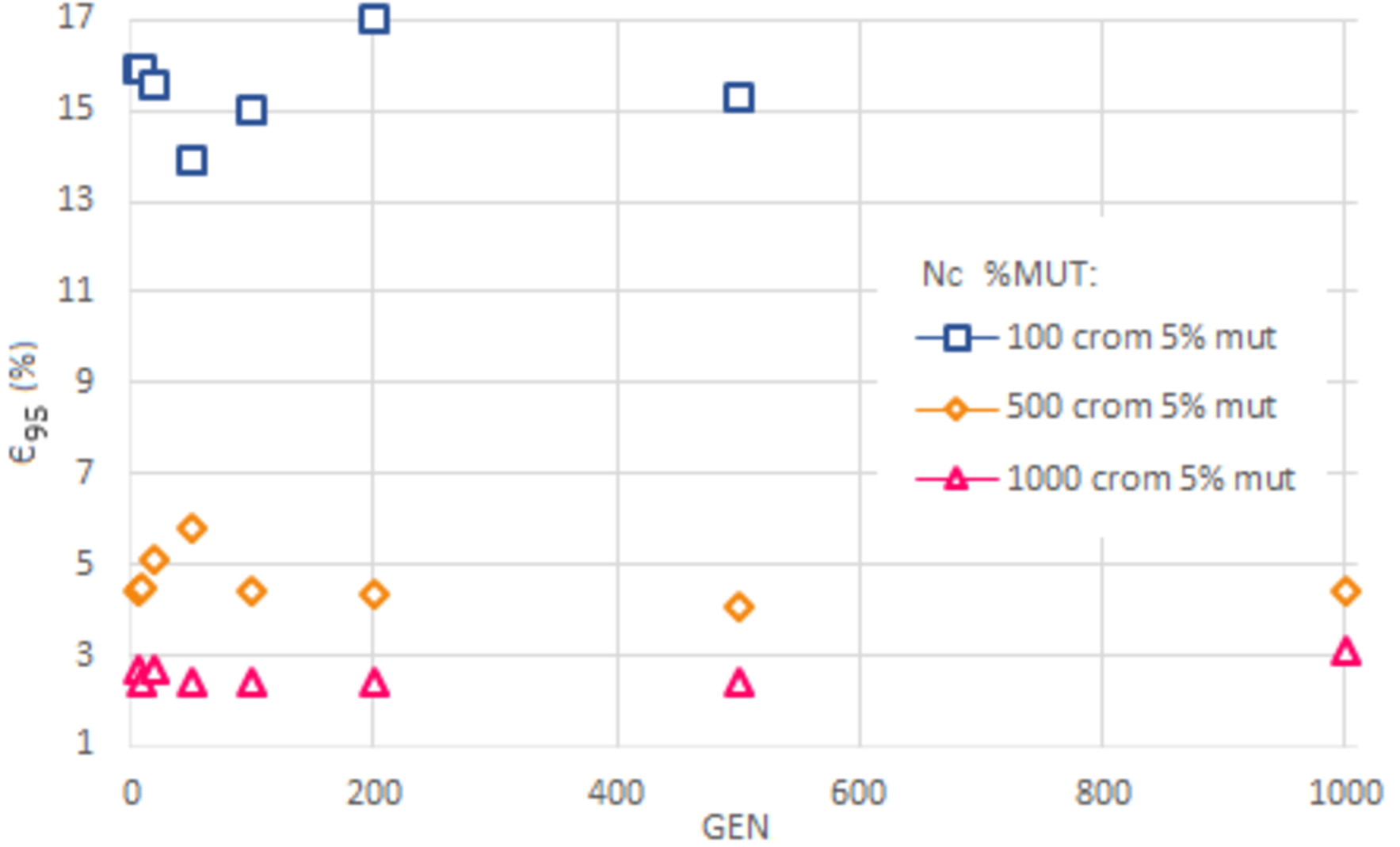}
\caption{$\epsilon_{95}$ as a function of the number of generations, $GEN$.  Each plot corresponds to $1000$ GA experiments with different numbers of chromosomes ($N_C$), and different percentages of mutations ($MUT$).}
\label{fig:E95_gen}
\end{figure}
In Fig. \ref{fig:E95_gen}, we confirm that $\epsilon_{95}$ is approximately constant whith $GEN$, for different $N_C$ values. As indicated in rows 4 to 6 of Table \ref{tab2}, $Eval_{GA}$ and $t$ increase significantly with $GEN$. 

Similarly, varying $MUT$ from $0$ to $20\%$ does not alter $\epsilon_{95}$ appreciably (Fig. \ref{fig:GA_acum} (c) and rows 7 to 9 of Table \ref{tab2}). 

\begin{figure}[ht]
\centering
\includegraphics[width=1\linewidth]{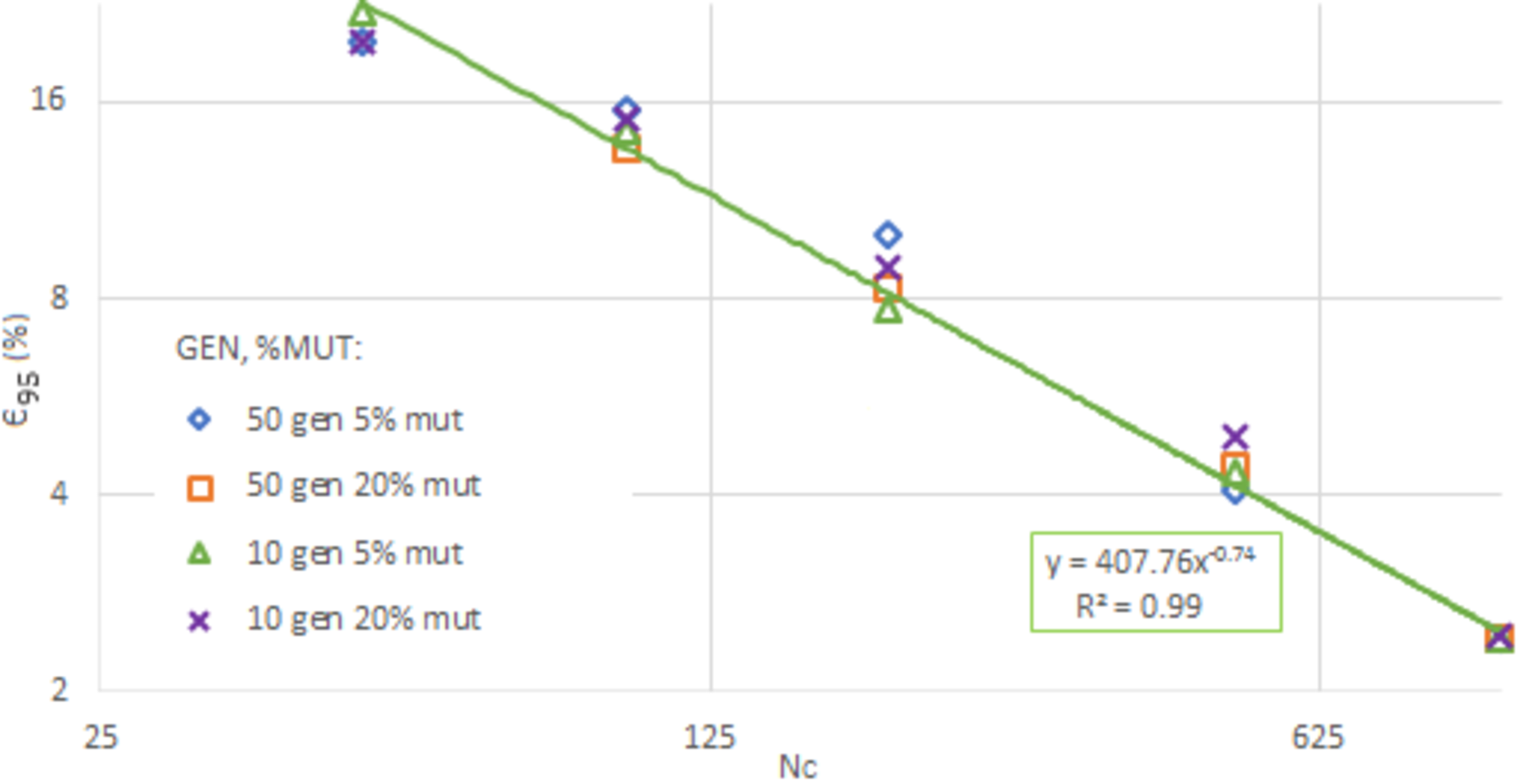}
\caption{$\epsilon_{95}$ as a function of the number of chromosomes, $N_C$. Each plot corresponds to GA experiments with different numbers of generations ($GEN$), and different percentages of mutations ($MUT$).}
\label{fig:E95_Nc}
\end{figure}
As shown in the log-log graph of Fig. \ref{fig:E95_Nc}, $\epsilon_{95}$ follows a power law with $N_C$, that is
\begin{equation}
\epsilon_{95} \propto N_C^{-\beta}, 
\label{eq14}
\end{equation}with $\beta \approx 0.74$, independently of $GEN$ and $MUT$. 

In summary, when using the Genetic Algorithm to optimise the design of photodetectors with three design parameters, it is best to select a large $N_C$ value, for example, $1000$ chromosomes, and a low $GEN$ value, like $10$ generations, since it leads to a lower $\epsilon_{95}$ and less computational cost. 

\begin{figure}[ht]
\centering
\includegraphics[width=0.95\linewidth]{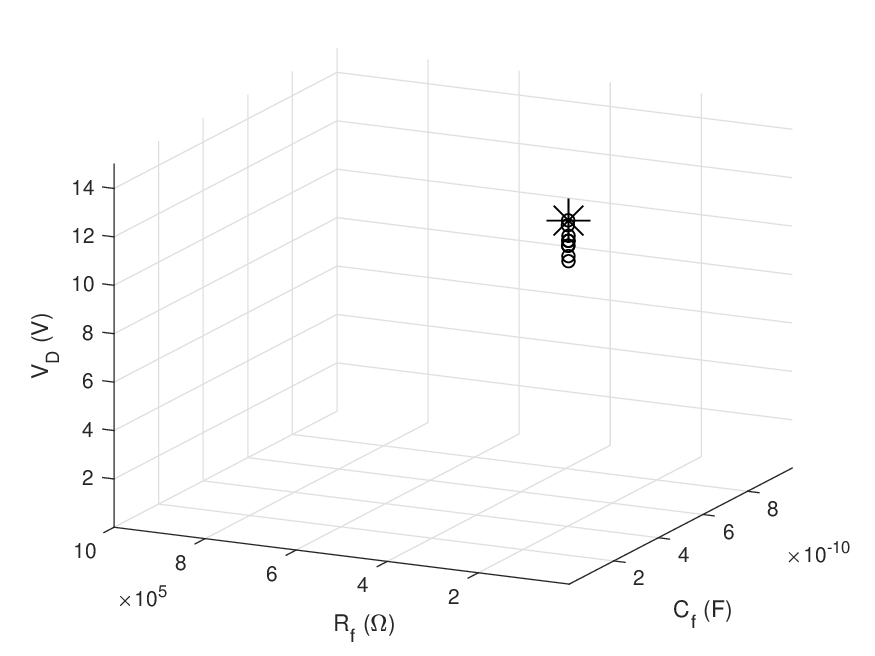}
\caption{Optimised chromosomes from the Genetic Algorithm experiment ($N_C = 1000$, $GEN = 10$, $MUT = 5\%$, and $\epsilon_{95} = 2.38\%$), that lead to circuits with $\epsilon \leq \epsilon_{95}$. The result marked with an asterisk corresponds to the optimum from systematic search.}
\label{fig:dispersionGA2-38}
\end{figure}
We compare the results of the systematic search and the Genetic Algorithm with a graphical representation in the design parameters space ($R_f$, $C_f$ and $V_D$). The numerical parameters of the GA experiment, in Fig. \ref{fig:dispersionGA2-38}, correspond to the fourth row of Table \ref{tab2} ($N_C = 1000$, $GEN = 10$, $MUT = 5\%$, and $\epsilon_{95} = 2.38\%$). In the figure, we plot with circles the circuits with $\epsilon \leq \epsilon_{95}$ (i.e. $Merit \geq 0.9116$). We can see that they are all concentrated in a small region of the plot. Also, in the figure an asterisk indicates the optimum circuit from the systematic search algorithm ($Merit_{syst} = 0.9338$). 

From the comparison of the Montecarlo and Genetic Algorithm results (Tables \ref{tab1} and \ref{tab2}) with the same number of evaluations ($10 \times 10^3$), we see that GA with $N_C = 1000$ has a much lower $\epsilon_{95}$ than MC ($2.38\%$ vs. $10\%$). However, the GA takes longer than MC ($8.2$ s vs. $0.72$ s). It must be remarked that both, Montecarlo and Genetic Algorithm, have a significantly lower computational cost than the systematic search ($10 \times 10^3$ vs. $3.7 \times 10^5$ evaluations).

\section{CONCLUSIONS}\label{Section_Conclusions}
In this work, we present Montecarlo and Genetic Algorithms for the optimisation of the design of photodetectors based on a transimpedance amplifier and a photodiode. The specifications are the signal-to-noise ratio, highest operating frequency, and phase margin. An additional requirement is to use commercial components values from the $5\%$ tolerance series. The design parameters are the components of the feedback network and the photodiode bias voltage. We define a merit function that allows a quantitative comparison of the compliance with the design requirements. The systematic search algorithm is used as a reference for the comparison of both optimisation algorithms. We  apply the optimisation algorithms to the same design example and compare their performances.  

The systematic search is used to find the design that has the highest possible merit. It requires the calculation of all the possible combinations of the design parameters values. The systematic search for the design example in this work requires approximately $3.7\times10^5$ evaluations. This method has the highest computational cost compared to the Montecarlo and Genetic Algorithm. 

To evaluate the performance of each optimisation algorithm, we define the percent relative difference, $\epsilon$, between the optimum merit (obtained by the systematic search) and the merit of the circuit optimised by the algorithm. It must be noted that $\epsilon$ is a random variable. The performance metric is given by $\epsilon_{95}$, that is the value of $\epsilon$ not exceeded in $95\%$ of the algorithm runs. 

For the Montecarlo algorithm with $10^4$ random sets of design parameters values, $\epsilon_{95}$ is $10\%$. The Montecarlo search has a significantly lower computational cost than the systematic search, since it gives good results even with about $3\%$ of the systematic search evaluations. Interestingly, $\epsilon_{95}$ follows a power law, decreasing with exponent $0.50$ with the number of random sets of design parameters values.

The Genetic Algorithm is based on the evolution of a number of random sets of the design parameters values (chromosomes) during a given number of generations and subject to mutations. This algorithm requires a total amount of computational evaluations that is the product of the number of chromosomes and the number of generations. Increasing the number of chromosomes lowers $\epsilon_{95}$, improving the performance of the Genetic Algorithm. It must be remarked that a large number of generations increases the computational cost without significantly improving $\epsilon_{95}$. In this work, we find that the percentage of mutations does not influence the results. Remarkably, $\epsilon_{95}$ for the Genetic Algorithm also follows a decreasing power law with exponent $0.74$ with the number of chromosomes.

We simulate the design example of this work through the evolution of $1000$ chromosomes during 10 generations ($5\%$ mutations). In this case, the Genetic Algorithm requires $10000$ evaluations resulting in $\epsilon_{95}$ of $2.38\%$. With the same amount of evaluations, the Montecarlo experiment results in $\epsilon_{95}$ of $10\%$. Even though for the same number of evaluations the Genetic Algorithm takes more computation time than Montecarlo, the $\epsilon_{95}$ is significantly lower. Since the power law exponent of the Genetic Algorithm is larger than that of Montecarlo ($0.74$ vs. $0.50$), to obtain the same $\epsilon_{95}$ ($2.38\%$), would take $176000$ evaluations for the Montecarlo algorithm.

We conclude that the merit function together with the optimisations based on Genetic Algorithm or Montecarlo are useful, robust and efficient tools for the design of analog circuits. It must be remarked that the higher power law exponent of the Genetic Algorithm is an advantage for its application to more challenging design problems with more parameters or constraints.

\section{ACKNOWLEDGMENT}
This work was supported by grants from Universidad de Buenos Aires (UBACyT 20020190100275BA and 20020190100032BA), CONICET (11220200102112CO) and by ANPCyT (PICT-2020-SERIEA-03741). One
of the authors (Patricia M. E. V\'azquez) thanks Facultad de Ingenier\'ia de la Universidad de Buenos Aires for a Peruilh FIUBA Doctoral scholarship.

\vfill

\end{document}